Title: The Making of a Genius: Richard P. Feynman

Author: Christian Forstner
Ernst-Haeckel-Haus
Friedrich-Schiller-Universität Jena
Berggasse 7
D-07743 Jena
Germany
Fax: +49 3641 949 502
Email: Christian.Forstner@uni-jena.de

Abstract:

In 1965 the Nobel Foundation honored Sin-Itiro Tomonaga, Julian Schwinger, and Richard Feynman for their fundamental work in quantum electrodynamics and the consequences for the physics of elementary particles. In contrast to both of his colleagues only Richard Feynman appeared as a genius before the public. In his autobiographies he managed to connect his behavior, which contradicted several social and scientific norms, with the American myth of the "practical man". This connection led to the image of a common American with extraordinary scientific abilities and contributed extensively to enhance the image of Feynman as genius in the public opinion. Is this image resulting from Feynman's autobiographies in accordance with historical facts? This question is the starting point for a deeper historical analysis that tries to put Feynman and his actions back into historical context. The image of a "genius" appears then as a construct resulting from the public reception of brilliant scientific research.

## Introduction

Richard Feynman is "half genius and half buffoon", his colleague Freeman Dyson wrote in a letter to his parents in 1947 shortly after having met Feynman for the first time.[1] It was precisely this combination of outstanding scientist of great talent and seeming clown that was conducive to allowing Feynman to appear as a genius amongst the American public. Between Feynman's image as a genius, which was created significantly through the representation of Feynman in his autobiographical writings, and the historical perspective on his earlier career as a young aspiring physicist, a discrepancy exists that has not been observed in prior biographical literature.

This discrepancy makes it necessary to highlight the heretofore released literature about Feynman in an introductory overview. The following analysis is based on Ludwik Fleck's studies of thought styles and thought collectives that are presented in the context of the American physical community and the person of Feynman. In the following passage, I will demonstrate this pragmatic thought style as the originating point of Feynman's ideas. To illustrate the gap between Feynman's own autobiographical depiction and the historical analysis, I will begin by examining initially from Feynman's foothold in the community of US physicists until the bestowal of the Nobel Prize in the year of 1965. The genius cult surrounding Feynman, which was created after the award ceremony, constitutes the last part of my portrayal. Thereby, I will contrast Feynman's autobiographical narrative against the historical analysis to show that Feynman's image as a genius came about due to the interaction of Feynman's character with the general public.

## Feynman's image in Literature

Feynman's character is the subject of three biographies: a scientific biography by physicist and historian of science Jagdish Mehra[2], a popular science but historically based biography by the scientific journalist James Gleick[3], and a popular biography by specialized book authors John and Marry Gribbin.[4] Of the three listed biographies, Gleick utilizes the most source material. Moreover, his biography is the only one that, at the least, critically questions Feynman's image as a genius at a rudimentary level. Mehra und Gribbin & Gribbin largely assume the image without criticism. Specifically, the latter simply skims the surface and can only conditionally be recommended for reading to interested non-professionals. Mehra presents a good and detailed introduction to Feynman's physical work. Gleick especially attempts to contextualize Feynman's physics. All three works are oriented around Charles Weiner's oral history interview with Feynman, which was commissioned by the American Institute of Physics (AIP).[5] Mehra conducted several interviews with Feynman shortly before his death, but the cited contents barely differ from that of the AIP. Particularly, Mehra's partially page long sequences of quotes from his interviews, without analyzing them critically, appear to be a careless handling of the historical sources.

---

[1] Richard P. Feynman [Jeffrey Robbins ed.], The Pleasure of Finding Things out: The Best Short Works of Richard P. Feynman, Foreword by Freeman Dyson (Cambridge, MA: Perseus Books, 1999), p. xi.

[2] Jagdish Mehra, The Beat of a Different Drum: The Life and Science of Richard Feynman (Oxford: Oxford University Press, 1996).

Likewise, physics historian Silvan S. Schweber conducted his own interviews with Feynman and analyzed Feynman's way to quantum electrodynamics in a comprehensive essay.[6] This essay was entered into the 8th chapter of Schweber's book formidable on the history of quantum electrodynamics.[7] In the book, Schweber gives an overview about the early background of the quantum field theory and quantum electrodynamics from the late 1920's through the war until the postwar era. Central to Schweber's book are the biographical studies of Sin-Itiro Tomonaga, Julian Schwinger, Richard Feynman und Freeman Dyson, in which he describes the genesis of modern quantum electrodynamics. But Schweber also assumed Feynman's anecdotes from the interviews largely without criticism.

---

[3] James Gleick, Richard Feynman: Leben und Werk des genialen Physikers (München: Droemer Knaur, 1993).

[4] John and Mary Gribbin, Richard Feynman: Die Biographie eines Genies (München, Zürich: Piper, 2000).

[5] Interview with Richard Feynman conducted by Charles Weiner, March 4 and 5, June 27 and 28, 1966, and Februar 4, 1973, American Institute of Physics, Center for the History of Physics, Niels Bohr Library, One Physics Ellipse, College Park, MD 20740, USA.

[6] Silvan S. Schweber, "Feynman and the Visualisation of Space-Time Processes," Review of Modern Physics, 58 (1986), pp. 449–508.

[7] Silvan S. Schweber, QED and the Men Who Made It: Dyson, Feynman, Schwinger and Tomonaga (Princeton: Princeton University Press, 1994).

In fact, the introductory overview of the existing Feynman biographies makes the necessity for a critical discussion between Feynman's image as a genius and the background of his biographies clear. Thereby, it is essential to historicize and contextualize Feynman's contribution to his image as genius through his autobiographical writings, interviews, and anecdotes. Thus, how closely Feynman's character as a scientist, his theories, and their genesis were interwoven with the thought style of his social community and the American physical community should be discussed next within the scope of a biographical analysis. In addition, from this background will be analyzed how Feynman contributed to his image as a genius in his autobiographical writings. The Feynman papers at the California Institute of Technology as well as the series of oral history interviews by the American Institute of Physics serve as source material.

## *Feynman and the Community of American Physicists*

In June 1947 an exclusive circle of just under 30 leading physicists from the United States of America met at Shelter Island, a small island outside of New York, amongst them John von Neumann, J. Robert Oppenheimer, Isidor I. Rabi und John A. Wheeler. In a casual atmosphere, they discussed the newest discoveries in atomic physics and quantum electrodynamics. Both of these sub domains of physics were implicitly determined as the two main fields of research for the postwar physics. The Shelter Island Conference and the following conferences in Pocono and Oldstone were therewith the most significant conferences for the development of physics after the Second World War.[8]

Photograph of the participants of the Shelter Island Conference

Richard Feynman, who didn't capture a central role at this conference in contrast to later conferences, stands hidden at the edge of the participant photo. According to the terminology of the polish doctor and scientific sociologist Ludwik Fleck, in this photography we see pictured a thought collective, with Feynman as part of this thought collective.

---

[8] Silvan S. Schweber, "Shelter Island, Pocono, and Oldstone. The Emergence of American Quantum Electrodynamics after World War II," Osiris 2 (1986), pp. 265–302.

In his essay *The Genesis and Development of a Scientific Fact*[9] released in 1935, Fleck develops, through a case study of the syphilis concept and the Wassermann reaction, his epistemic and scientific sociological conclusions with the central terminology of the thought collective and thought style. Science recognized Fleck as social action and generally identified a community of people whose mental exchange and thoughts were interdependent on another as a thought collective. This is the means to the historic development of a thought domain, a certain knowledge and cultural inventory, a thought style that offers the theoretical conditions upon which the group builds their knowledge.

The participant photo of the conference shows such a thought collective, more precisely through the esoteric circle of the more comprehensive who only in part represent this thought collective of US physicists. This esoteric circle establishes elite within the thought collective that decisively shapes the development of the field. The participants at the Shelter Island Conference did this as well and during the conference, implied both the main fields of research for postwar physics of the USA: quantum electrodynamics and particle physics. The rules for thought and collaboration of such a group embody their thought style. The thought style can also be understood as a directed perception, as coherent perception with corresponding mental and objective processing of the perceived. It is characterized by collective traits of the problems and judgments that are seen as evident, the methods that are applied as a means to recognition, and the determining of which questions appear to be nonsensical.

Feynman, who received the Nobel Prize in physics in 1965, developed his own concept, not as part of this collective, but rather seemingly disassociated from all social ties as a genius context free of his physical works. This image of Feynman formed and was significantly shaped by his autobiographical narratives and anecdotes that stem from the time after Feynman's receipt of the Nobel Prize. In the course of this production, Feynman appears to belong to no thought collective. Therefore, this study centers on the historically critical argument with Feynman's image as a genius and the attempt to collectivize his works, meaning to understand these from the thought collective, which as a matter of fact, Feynman never left.

---

[9] Ludwik Fleck, Entstehung und Entwicklung einer wissenschaftlichen Tatsache: Einführung in die Lehre vom Denkstil und Denkkollektiv (Frankfurt am Main: Suhrkamp, 1980).

## *The Style of US Physics: Practical Men und Cultivated Persons*

In the 19th century the spirit of the ''practical man'', who made use of natural science to be able to better acquire natural resources, predominated at public colleges. At private colleges, liberal arts determined the canon of the student's education. Natural sciences were frequently delegated or separated to technical schools, such as at Harvard the Lawrence School of Science or at Yale the Sheffield School. At the beginning of the 1870's, a turning point set into public awareness. One who could participate in discourse about natural science was regarded as a ''cultivated person''. As a result, many new private universities were founded such as Cornell and John Hopkins University, which followed a clear research imperative. The private investors that had acquired their assets in business and industry were regarded as ''practical men'' and through the funding of pure natural sciences, enhanced their status amongst the public's awareness. By means of funding abstract natural sciences, they gained the status as a ''cultivated person''.[10]

Next to the development of a clear research imperative, a tradition of precision measurement had emerged in the USA in the 19th century that went above and beyond a simple gathering of facts.[11] One example is the failed attempt to verify ether by Albert A. Michelson and Edward W. Morley. In contrast, theoretical physics existed only rudimentarily and was represented by individual people without exhibiting a larger structural context. In the course of the reception of the Bohr-Sommerfeld quantum theory, the necessity for the development of this theory was recognized. Leading European ambassadors of the theory were invited to lecture tours through the US, and postgraduates and post doctoral students were delegated to centers in Europe and then established new schools of theoretical physics in the USA. The European research style within the scope of this transfer was not assumed 1:1. In fact, the theory was integrated into a new thought collective and adjusted to the thought style of US physicists. When theoretical physics in the USA emerged out of the shadowy existence of experimental physics at the end of the 1920's, a close connection was maintained between both branches. As a rule, every physicist had to pass through an experimental basic education and research courses were carried out collectively. Interest in an experimental inspection of computations went so far, that it was seen as the duty of the theorist, not to engage in any philosophical speculation, rather to exclusively carry out calculations whose experimental verifiability were guaranteed. This pragmatic basis led to it that US physicists reflected on the philosophical aspects of their work much less frequently than in Germany.[12] Exemplary for this is an excerpt from the theoretical physicist John C. Slater quoted from the year 1938, in which he elevates the development of experimental prognoses to the main task of a theorist:

> „A theoretical physicist in these days asks just one thing of his theories: if he uses them to calculate the outcome of an experiment, the theoretical prediction must agree within limits, with the results of the experiment. He does not ordinarily argue about philosophical implications of this theory. Almost his only recent contribution to philosophy is the operational idea, which is essentially only a different way of phrasing the statement I have just made, that the one and only thing to do with a theory is to predict the outcome of an experiment [...] Questions about a theory which do not affect its ability to predict experimental results correctly seem to me quibbles about words, rather than anything substantial, and I am quite content to leave such questions to those who derive some satisfaction from them."[13]

---

[10] see Daniel J Kevles, The Physicists: The History of a Scientific Community in Modern America (Cambridge, MA: Harvard University Press, 1995, 3rd edition), chapter I and II, pp. 3–24.

[11] Marie-Noëlle Bourguet, Christian Licoppe and H. Otto Sibum, (eds.), Instruments, Travel and Science: Itineraries of Precision from the Seventeenth to the Twentieth Century (London: Routledge, 2002).

In the postwar era, this pragmatic research style changed, which was identified by characters such as Slater who joined with the younger theoretical physicist, worked very closely together with experimental physicists, and knew their needs exactly. In particular, in particle physics and military research the work of experimenters and theorists merged together so much, that the old pragmatic style that labeled the generation of Slater, further developed into a ''pragmatic, utilitarian, instrumental style''.

---

[12] Silvan S Schweber, "The Empiricist Temper Regnant: Theoretical Physics in the United States 1920–1950," Historical Studies in the Physical and Biological Sciences, 17 (1986), pp. 55-98, and Kevles, Physicists (ref. 10), chapter I–VI, pp. 4–90; as well as Paul Forman, "Social Niche and Self- Image of the American Physicist," in Michelangelo de Maria, Mario Grilli, and Fabio Sebastiani eds., The Restructuring of Physical Sciences in Europe and the United States 1945–1960 (Singapore, New Jersey, 1989), pp. 96–104, here p. 99-100.

[13] John C. Slater, "Electrodynamics of Ponderable Bodies," Journal of the Franklin Institute 225 (1938), pp. 277–287, here p. 277.

## *Feynman and the Thought Collective of US Physics*

Scientific education does not only aim to impart student knowledge and work techniques, but also to mediate a certain method of thought and research, a thought style, to potential young academics. This initiation process is accompanied by a pressure on the individual to adapt in the form of assignments, tests, etc, until, after successful knowledge acquisition and adaption, the individual is accepted into the scientific community through a variety of initiation rituals.[14]

The young Feynman appeared to have no problems with the integration mechanisms into the collective. On the contrary: he passed the adaption measures so quickly, that he still had time and space leftover for his private studies. Feynman's adaption of the thought style can already be traced during his school days. Outside of regular class, he dedicated considerable time to his private studies and quickly took a pragmatic stance, which showed that Feynman attached only a slight meaning to the approach, as long as the outcome was correct.[15] Or according to Feynman's words: "I wouldn't give a damn – I know I didn't care – to find out what way you had to do it, because it seemed to me, if I did it, I did it."[16]

In the scope of his private studies, Feynman didn't removed himself from the thought style of the US American physical community,[17] rather assimilated their way of thinking outside of the designated mechanisms. This also becomes clear during his undergraduate studies at Massachusetts Institute of Technology (MIT), where he enrolled in mathematics, then followed the American myth of the practical man, switched to engineering sciences, before he found his ultimate major in physics. The so-called myth also showed up in Feynman's physical works, when he rhetorically placed experiment above theory. Feynman was introduced to theoretical physics through classic American textbooks. The textbook by John

---

[14] Fleck, Wissenschaftliche Tatsache (ref. 9).

[15] Interview Feynman-Weiner, March 4, 1966 (ref. 5), pp. 16-53. See also Gleick, Richard Feynman (ref. 3), pp. 55-60, and Schweber, QED (ref. 7), p. 374, as well as Mehra, Different Drum (ref. 2), chapter 2, pp. 22–43.

[16] Interview Feynman-Weiner, March 4, 1966, p 27.

[17] Christian Forstner, Quantenmechanik im Kalten Krieg (Diepholz, Berlin: GNT, 2007), pp. 59–76, and Schweber, The Empiricists (ref. 12), pp. 55–98 and for the postwar era Silvan S. Schweber, „Some Reflections on the History of Particle Physics in the 1950s," in Laurie M. Brown, Max Dresden and Lilian Hoddeson, eds., Pions to Quarks: Particle Physics in the 1950 (Cambridge: Cambridge University Press, 1989), pp. 668–693.

C. Slater and Nathaniel H. Frank, [18] which the courses at MIT were based on, had the professed goal to educate the students as productive physicists. In the American understanding of a theoretical physicist, productive meant that this measurable prediction fits without engaging in philosophical speculation. The numerous assignments at MIT didn't present Feynman with any problems. He executed them, adopted the thought style-appropriate knowledge, and was integrated into the collective in this way. Together with a fellow student, he took up private studies. They laid aside the abstract textbook by English physicist Paul A.M. Dirac, and instead, chose the textbook of Americans Linus Pauling and E. Bright Wilson, [19] which was oriented on the practical use of quantum mechanics.

Independently, they derived the Klein-Gordon equation, but, with the application on the hydrogen atom, had to conclude that the practical test of the equation failed. Feynman later called this experience a defining moment, in which he recognized that the beauty of an equation was not the decisive point of physics, rather the test of the predicted results against the experiment. Once again, we can detect Feynman's reference to the thought style of American physics. In his paper ''Forces Molecules'', Feynman detached himself from the standard observation about energy minima and directly calculated the working forces. Thereby, he concluded that according to the quantum mechanical calculation of charge distribution only the classical electrostatic forces appeared in the equations. [20] In a direct way, Feynman found a method to link time saving and higher accuracy of the results. [21] As a graduate student at Princeton, Feynman, assistant to the theoretical physicist John A. Wheeler, continued to work on the fundamental problems of quantum electrodynamics. Feynman expressed himself enthusiastically, to finally have arrived at the frontline of current research, meaning he was thrilled by the questions that the thought collective and thought style specified. This is a further indication of Feynman's integration into the thought style. Feynman's position that the way to the goal didn't matter to him, as long as the experimental values were consistent, appeared at numerous opportunities at Princeton as well. Together with Wheeler, Feynman researched the scattering matrix theory and pair production. Independently, Feynman once again started research ideas on the endless self energy of the electron, which he had drawn from one of Dirac's books in his undergraduate studies at MIT. First of all, he tried to get a handle on the problem through the elimination of the field concept in classical theory. For this purpose, Feynman and Wheeler regarded the entire space-time path that a particle chooses. The emphasis is on ''chooses'' because according to the principle of least action that both of them adopted, a particle teleological ''chooses'' a path, on which the least amount of energy is transformed. The classical theory appeared to function well, but heeler and Feynman broke down the expansion of the theory to quantum electrodynamics. [22]

---

[18] John C. Slater and Nathaniel H. Frank, Introduction to Theoretical Physics (New York: McGraw-Hill, 1933).

[19] Linus Pauling and E. Bright Wilson, Introduction to Quantum Mechanics: With Applications to Chemistry (New York, McGraw-Hill, 1935).

[20] Richard P. Feynman, Forces and Stresses in Molecules, 1939, Bachelor Thesis, unpublished, MIT Archives.

[21] Forstner, Kalter Krieg (ref. 17), pp. 155–159.

Already incorporated into the first war research, Feynman summarized the results of these works in his dissertation and arrived at the first application of the action principle in quantum mechanics. Feynman described the path to this point more than 20 years later in a type of dream story. During the day, he had worked with the German emigrant Herbert Jehle on an article of Dirac, and during the night, Feynman then lay in bed and suddenly saw the solution appear in front of him:

> "I was lying in bed thinking about this thing, and thought, "What would happen if I wanted to get the wave function at one time, and at finite interval later suppose that the interval was divided into a large number of small steps?… I could represent the coordinates that I was integrating over a succession of positions through which the particle was supposed to go, and then this quantity, this sum, would be like an integral the integral of L, which is in fact the action… I saw the action expression, suddenly, so to speak… In the air, in the head. Yeah. You see the action coming on. And I said, "My God, that's the action! Wow!" I was very excited. So I had filed a new formulation of quantum mechanics in terms of action, directly. I got up and wrote everything out, and checked back and forth, and made sure it was all right, and so on."[23]

Feynman's story of the dreams of the paths is also in his Nobel Prize speech. This story has both the function of emphasizing his intuition, and at the same time brings about a break from tradition.[24] This narration is strongly reminiscent of Friedrich A. Kekulés' famous daydream about the discovery of the benzene ring, of which he similarly, after more than 20 years after the discovery, reported:[25]

---

[22]Ibid., pp.159–168.

[23]Interview Feynman-Weiner, March 5, 1966, p. 155.

> I turned the chair toward the fireplace and dozed off. Once again, the atoms were mocking me in front of my eyes. This time, smaller groups lingered modestly in the background. My mind's eye, sharpened through repeated faces of similar types, was now able to distinguish larger structures from manifold configuration. Long rows, multiple times more densely joined together, everything in movement, curling and turning in a snake-like manner. And look, what was that? One of the snakes grasped its own tail and tauntingly whirled the structure before my eyes. As if being awakened by a lightning bolt I woke up; once again, I spent the rest of the night working out the consequences of the hypothesis.[26]

Different than the descriptive ''holy apparition'' of the path integral formulation presumes, Feynman purposefully tied in the prior works of the community into his dissertation and emphasized the usefulness of his research for the current quantum electrodynamics and particle physics research programs. Similarly, according to style, he pointed out that the experimental testability of the theories was significant criteria for their accuracy:

> "The final test of any physical theory lies, of course, in experiment. No comparison has been made in this paper. The author hopes to apply these methods to quantum electrodynamics. It is only out of some such direct application that an experimental comparison can be made."[27]

---

[24]Interview Feynman-Weiner, March. 5, 1966, p. 155, and Richard P. Feynman, "The Development of the Space-Time View of Quantum Electrodynamics: Nobel Lecture, December 11, 1965," in Nobel Lectures: Physics, 1963–1970 (Amsterdam: Norstedt, 1972), pp. 155–178, here pp. 166-167. See also the uncritical reception at Mehra, Different Drm (ref. 2), p. 137.

[25] A. J Rocke, "Hypothesis and Experiment in Kekulé's Benzene Theory," Annals of Science, 42 (1985), pp. 355–381.

[26] Gustav Schultz, „Bericht über die Feier der Deutschen Chemischen Gesellschaft zu Ehren August Kekulés" Berichte der Deutschen Chemischen Gesellschaft, 23 (1890), pp. 1265–1312, hier quoted from Kekulés' talk, p. 1306.

During the Second World War, Feynman worked in Los Alamos as the leader of a theoretical research group on the central project of the construction of the atomic bomb. The war research in Los Alamos was characterized by the specific work style of the scientists. Instead of working on single fundamental theoretical problems, they worked in an application-oriented manner taking small steps towards a solution to the problem. Similarly, the theorists worked closely together with experimental physicists and engineers. In place of exact solutions, general estimations of the results sufficed within clearly defined tolerances, and well-defined work instructions made these results usable for further groups of people. This work style also shaped Feynman's personal style in the immediate postwar period, which newly clarified his openness and receptiveness for the predetermined guidelines of a social community. Furthermore, Los Alamos functioned as a career network, and after 1945, Feynman was nearly overwhelmed by job offers. This newly clarified his strong foothold in the community. [28]

For Feynman, the theorist and German emigrant Hans Bethe was a central attachment person, who promoted him in the postwar period. Still in Los Alamos, Bethe had successfully enlisted Feynman as a professor at Cornell University. Bethe sponsored Feynman, supported him at diets and established contact between Feynman and the leading-edge of the physical community. After 1945, Feynman got involved with the civil control of atom research through public lectures in the scope of the Federation of Atomic Scientists. [29] This is contrary to the image of the genius who is detached from all social, which Feynman illustrates of himself in his later autobiographical writings.

---

[27]Richard P. Feynman, The Principle of Least Action in Quantum Mechanics (Princeton: PhD-Thesis, Princeton University, 1942).

[28] Peter Galison, "Feynman's War. Modelling Weapons, Modelling Nature," Studies in the History and Philosophy of Modern Physics 29 (1998), pp. 391–434. See also Gleick, Richard Feynman (ref. 3), pp. 227–300 and especially Schweber, QED (ref. 7), pp. 397–405.

[29]FBI-File of Richard Feynman, received under the Freedom of Information Act, Memorandum [ca. 1946, not dated]. See also: Interview Feynman-Weiner, June 27, 1966, pp. 68-69. Feynmans postwar engagement is shortly mentioned in Schweber, QED (ref. 7), pp. 469-470 and Mehra, Different Drum (ref. 2), p. 171 mentions one talk of Feynman without putting it in the context of the Federation of Atomic Scientists.

We can discover a different image of genius in Feynman's story about his quantum electrodynamics research: As a young physicist being under high expectations and competitive strain, Feynman suffered a burn-out at Cornell and decided to only do that, which was fun for him. In his later autobiographical narrative, after this decision he arrived at his version of quantum electrodynamics through the just-for-fun analysis of the precession movement of a plate:

> It was effortless. It was easy to play with these things. It was like uncorking a bottle: Everything flowed out effortlessly. I almost tried to resist it! There was no importance to what I was doing, but ultimately there was. The diagrams and the whole business that I got the Noble Prize for came from that piddling around with the wobbling plate.[30]

The path from the rotating plate to quantum electrodynamics remains a secret to the non-professional. It is no longer traceable and similar to the daydream about the path integral formulation, appears utterly ''ingenious''. As a matter of fact, Feynman's work about his version of quantum electrodynamics consisted of a constant process of translation between Feynman's language and that of the collective. Bethe accompanied Feynman during this translation process and supported him. Feynman presented the modern version of the path integral formulation in 1947 at the Shelter Island Conference without meeting a great response from his colleagues. His formulations were too foreign for the participants at the conference. Additionally, he compiled a publication at the suggestion of a befriended colleague, which ultimately appear in the Review of Modern Physics in 1948.[31] The publication contained in an axiomatic formulation a version of the path integral, in which every possible virtual path is allocated amplitude and phase, so that every conceivable virtual path delivers a contribution to the cumulative feasible amplitude. Likewise, this publication formed through a tight discussion process between Feynman and members of the collective.[32]

---

[30] Richard P. Feynman, "Surely you're Joking, Mr. Feynman!" Adventures of a Curious Character (New York: Norton, 1997), p 174.

[31] Feynman, Richard P. (1984). Space-Time Approach to Non-Relativistic Quantum Mechanics. In: Reviews of Modern Physics, 20, S. 367–387.

[32] See Gleick, Richard Feynman (ref. 3), pp. S. 334–339, for Shelter Island and Mehra, Different Drum (ref. 2), p. 183 for the orgination of Feynman's paper, as well as Schweber QED (ref. 7), pp. 409–413 for Shelter Island and the paper.

After the Shelter Island Conference, Feynman continued to work on his formulation of quantum electrodynamics. Bethe supported him in the calculations of the Lamb-Shifts, a displacement of energy levels in hydrogen atoms, which can only be explained through effects of quantum electrodynamics. As his mentor, Bethe motivated Feynman to write publications and helped him with the translation of the results into the language of the community, as Feynman later remembers:

> So I made the corrections. I tried to translate it back in the language that other people use. And then I went in, the next day. He [Bethe] showed me how to calculate the self-energy of an electron, and I showed him what the correction ought to be. I tried to translate my principles into the other language that he was explaining this thing in.[33]

Feynman compiled his results while under significant competitive strain. His colleague Julian Schwinger always seemed to be a step ahead of him and in addition, was better positioned in the community due to his numerous publications. At the follow up conference of Shelter Island in Pocono, Feynman introduced his results on quantum electrodynamics. Feynman's unusual approach was met with surprise and Freeman Dyson's transfer of Feynman's approach into style consistent knowledge first allowed a wide spread reception of Feynman's theory. In 1965 Feynman, along with Schwinger and Tomonaga, received the Nobel Prize for his works on quantum electrodynamics.

After the bestowal of the Nobel Prize, a genius cult developed around Feynman. The development of this cult was strengthened through autobiographical anecdotes, which later appeared as autobiographical volumes, which will be analyzed in the following.[34]

## *Feynman as a Genius*

On the 21$^{st}$ of October, 1965, Feynman received a telegram from Stockholm that congratulated him, Julian Schwinger, and Sin-Itiro Tomonaga for obtaining the Nobel Prize. Feynman's life changed with the Nobel Prize: the young, promising physicists from Los Alamos, who competed with Schwinger in quantum electrodynamics, had ultimately become a public character and reached an unforeseen popularity as no American physicists had before him. Up to the present day, a cult has developed around Feynman's character. The drumming physicist Feynman adorns posters and book jackets. Recordings of his drum pieces are acquirable by purchase; his written works became bestsellers, both autobiographical and popular science as well as specialized science. Feynman-Diagrams decorate the church windows of the Saint Nikolai Church in Kalkar in Germany; Feynman pops up in comics. The cult around the 'genius Feynman' is solely surpassed by the cults around Albert Einstein, whose image, in the meantime, is available to buy in the form of Albert Einstein action figures and 1,60m tall standing cardboard figures.

---

[33] Interview Feynman-Weiner, June 27, 1966, p. 29.

[34] Forstner, Kalter Krieg (ref. 17), pp. 186–196.

In the preface already cited above, Freeman Dyson compared the relations of his character to Feynman with the English authors Johnson and Shakespeare. Jonson mastered his craft as an author, while Shakespeare distinguished himself through the 'genius'. In this comparison, Dyson identifies with Jonson and Feynman with the genius of Shakespeare.[35] This comparison followed Feynman's death by more than ten years and is the only one known to me, in which a scientist explicitly depicts the scientist of Feynman as a genius. Scientists depict one another as highly gifted or talented, but do not depict each other with the term genius. This term arises from the interdependency between scientists and the associated public in which they move. The term genius can be summed about in three different points:

The magician as an element of the genius image stands out due to the fact that the way to the scientific results is no longer comprehensible for the environment and that this magic formula is seemingly created free of context. The above cited dream stories of Feynman and Kekulé are fundamental to this element.

The genius image in Feynman's stories is further characterized by the fact that he seemingly detached himself from the norms of his scientific as well as social environment and through this was removed from the world.

The ''revolutionary'' is the last element of Feynman's image as a genius and contributes a new perspective of a subject area in the community and the public through research results.[36]

All three points are necessary to comprehend the 'genius Feynman'. For example, if the third point was missing, a crazy bat remains, but not a scientific genius. The first two points of the definition of a genius can be strengthened by the self-portrayal of scientists to the public, while in the third point the popular portrayal of natural science wins in meaning. In contrast to Schwinger and Tomonaga, Feynman only acquired the reputation of a genius amongst the public. In his biographical self-portrayal, he was successful in connecting social and scientific norms of a contradictory manner with the American myth of the 'practical man'. In the public, this presented him as a 'typical' American with uncommon scientific capabilities and significantly contributed to his crowning as a genius in the public's awareness. [37]

---

[35] Feynman, Finding Things Out (ref. 1), Foreword by Dyson, p. viii.

[36] The term "revolutionary" is not used in the Kuhnian meaning.

## *Feynman as a Magician*

The scientific genius as a magician ''sees'' solutions, while the typical scientists has to derive them from established laws. Feynman systematically built this image of a genius up in his most well-known autobiographical anecdote volume titled *Surely you're Joking Mr. Feynman.* [38] In it, he began with a story about his High School algebra club. The time for a solution to the posed assignments was too short for a conventional way to the solution. Because of this, it was necessary to arrive at a solution with the help of tricks or as even Feynman reports: ''Sometimes I could see it in a flash.'' [39] Later, during his graduate studies at Princeton, he discovered that ''monster minds'' such as John A. Wheeler, John von Neumann and Albert Einstein could see the solution, while he still had to tediously calculate them. [40] Later, when he reported about the meaning that scientific teachings had taken for him, he separated himself from the scientists at the Institute of Advanced Studies at Princeton. Because these scientists exclusively sat in offices to gather their thoughts, they did not receive inspiration and were no longer in a condition, to develop new ideas. [41] This argument of Feynman's brings about the result that the scientists, until now assessed as ''monster minds'', decrease in favor. In the following chapter, Feynman then refers to his own creativity that, among other things, led to his formulation of quantum mechanics through the precession movement of a plate. During a conference in Japan, Feynman ultimately saw the solution of physical problems for himself. [42] Therewith, the argumentation follows in three points: First, the establishment of the ''monster minds'', then their degradation based on their lack of creativity, and finally the assertion of his own creativity and physical intuition. With this, he strengthened the image of the exceptional personality and the ingenious scientist of Richard Feynman.

---

[37] Gleick, Richard Feynman (ref. 3), pp. 450–477, discusses the question "what is a genius?" with a strong focus on the genius as magician and his originality but disregards the role of the public in the process of making a genius.

[38] Feynman, Joking (ref. 30).

[39] Ibid., p. 23.

[40] Ibid., p. 78.

[41] Ibid., p. 165.

## *The Genius removed from the World*

The image of the genius removed from the world was depicted through the apparent detachment of the actor from his social context. This occurs in Feynman's stories through the thematic of breaching the norms on two levels: a scientific one and a social one.

In the autobiographical anecdote volume, *Surely you're Joking Mr. Feynman,* he presents such breaching of norms amply. Already the title of the book arises from such breaching of norms: Feynman had requested a tee with milk and lemon at the dean's teatime, to which he received the response ''surely you're joking, Mr. Feynman''. A genius does not request tee with milk and lemon, but Feynman's presentation of this small mishap clarifies that he made fun of the conventions of Princeton's academic elite. [43] This becomes even clearer, when he describes the language of Princeton's academics, which he characterizes as an imitation of the English colleges of Cambridge and Oxford. Here, Feynman distances himself from the conventions of a social group, to which he himself belonged and consciously separates himself from Princeton's upper class. [44]

In the social framework of science, Feynman repudiated the norms. Thus, he did not publish his results frequently, did not participate as a referee in the system of self-assessment of scientific journals and did not compose any review articles. Similarly, he rarely took part in faculty meetings. [45] Also, Feynman consistently rejected honorary doctorates and acceptance into scientific societies. Prizes and honors in science are to be understood as recognition for the work of the researcher, who ultimately leaves society his results and receives recognition in return. Initially, Feynman did not react to his appointment into the American Philosophical

---

[42] Ibid., pp. 244-245.

[43] Ibid., p. 60.

[44] Ibid., pp. 59-97.

[45] Interview Feynman-Weiner, June 28, 1966, pp. 131, 169-170.

Society. Not until he received an assertive request to get in touch, did he reject acceptance two and a half months. [46] Feynman's reason for the rejection of the membership was not due to a rejection of the respective society, rather a general rejection of the system of scientific recognition. Thus, he justified his withdrawal from the National Academy of Science with a letter to the President in the year of 1961:

> The care with which we select "those worthy of the honor" of joining the Academy feels to me like a form of self-praise. How can we say only the best must be allowed to join those who are already in, without loudly proclaiming to our inner selves that we who are in must be very good indeed? Of course I believe I am very good indeed, but that is a private matter and I cannot publicly admit that I do so, to such an extent that I have the nerve to decide that this man, or that, is not worthy of joining my elite club.[47]

Similarly in his anecdote volume, he picks out the central theme of breaching norms that stand up against the norms of civil society. To this count his regular visit to topless bars, the drawing up of a painting for a brothel, and the testimony in favor of the topless bar's owner in court. Feynman not only rejected civil norms that to him seemed to be part of a double standard, but also publicly illustrates this rejection. [48] Similarly, Feynman's experiences with women and gambling in Las Vegas appeared as not compliant with a civil society. [49] The same is true of Feynman's description of hallucinations that he experienced in a special studio. This studio had water filled tanks available, in which one could allow oneself to be locked in, in order to experience hallucinations. [50] These experiences and Feynman's experimentation with testing his own limitations, allow for him to appear as a crazy bat to the average American.

---

[46] Letter from George W. Corner to Richard Feynman, April 19, 1968, Richard Phillips Feynman Papers (RPFP), California Institute of Technology, Institute Archives, Mail Code 015A–74 Pasadena, CA 91125, USA, Folder 1.1. and Feynman's refusal July 1, 1968.

[47] Letter from Richard Feynman to Detlev W. Bronk, August 10, 1961, RFPF, Folder 1.13.

[48] Feynman, Joking (ref. 30), pp. 270–275.

[49] Ibid., pp. 221–231.

[50] Ibid., pp. 335-337.

In spite of his continual breaching of norms that affected not only the civil American public, but also the social framework of science, Feynman did not experience any kind of sanctions. Since Los Alamos, Feynman was strongly anchored in the scientific thought collective. This clarifies, for instance, the numerous job offers that he received after 1945, as well as his participation at exclusive postwar conferences, such as the Shelter Island Conference and the Pocono Conference.[51] The bestowal of the Nobel Prize to Feynman also depicts the high regard that the community showed Feynman. Furthermore, with his approach, Feynman significantly contributed to the solution of the problem of quantum electrodynamics. His work style always remained within the frame of the thought style of the community and did not question their systematic basic principles. In spite of his out of the ordinary behavior, Feynman offered numerous connecting points even to the public.

In his popular science writings, a rejection of the behavior of the educated civil elite goes along with the transfiguration of practical manual tasks and a rejection of philosophical works, when Feynman, for instance, determines: „In the early fifties I suffered a disease from the middle age: I used to give philosophical talks about science."[52] Therewith, it indeed becomes clear why Feynman was not outlawed due to his later breaching of norms, but in order to understand why he was praised as a physical genius, it is necessary to examine the genius as a revolutionary.

## *Genius as Revolutionary*

Feynman's formulation of quantum electrodynamics, similar to Schwinger and Tomonaga's operator formalism, was not revolutionary in terms of Kuhn's paradigm shift. A break from previous physics did not follow, rather the new theory established ties to quantum mechanics.[53] The revolution, understood as a central innovation in Feynman's approach, was part of a different domain, which his former competitor Julian Schwinger characterizes in the following: „Like the silicon chip of more recent years, the Feynman diagram was bringing computation to the masses."[54] Feynman's way enabled physicists to arrive at results in a quick way, and in comparison to Schwinger and Tomonaga's, without complex calculations.[55] A colleague of Feynman's already determined this after receiving an advanced copy:

> I want to thank you for forwarding prepublication copies of your papers on electrodynamics. I have found them very interesting and have tried applying your methods not yet (except in an exploratory way) in new problems. I also gave a series of three or four seminars on them. They certainly are wonderful in making calculations easier.[56]

---

[51] For the postwar conferences see Schweber, Shelter Island (ref. 8), pp. 265–302.

[52] Ibid., p. 279.

[53] Helge Kragh, Quantum Generations: A History of Physics in the Twentieth Century (Princeton: Princeton University Press), pp. 332-336.

Through the translation of highly complex mathematical theory into simple language, tied in with a graphic viewing apparatus, the popularization of this theory was made possible. Therein consists the second ''revolution'' of Feynman's works. Therefore, not only did he make quantum electrodynamics more wide spread among the physical community, but also brought popularity to modern physics, like very few other physicists of the 20$^{th}$ century, through his numerous lectures that were later publicized as essays and in books.

*The Character of Physical Law*[57] was created from a series of lectures that Feynman had held in 1960. In this volume, Feynman leads the reader to complex and fundamental questions from Newton's law of gravitation to the concept of symmetry in modern physics, which he visualizes without sliding into the trivial. In the process, Feynman also ventures on to complex themes. The volume *QED -The Strange Theory of Light and Matter* similarly emerged from a popular lecture series and, to this day, is a unique popular scientific introduction to Feynman's formulation of quantum electrodynamics.[58] His capability to make such complex data accessible to others, made Feynman famous as a teacher and textbook author. The famous Feynman Lectures emerged from a multiple semester lecture series and fortified Feynman's reputation as a brilliant teacher. [59]

---

[54] Julian Schwinger, "Renormalization Theory of Quantum Electrodynamics," in: Laurie Brown and Lilian Hoddeson eds, The Birth of Particle Physics (Cambridge: Cambridge University Press, 1986), pp. 329–353, quoted from p. 343.

[55] For path integrals and their importance as a technique see David Kaiser, Drawing Theories Apart: The Dispersion of Feynman Diagrams in Post-War Physics (Chicago, University of Chicago Press, 2005).

[56] Letter from Robert F. Christy to Richard Feynman, September 12, 1949, RPFP, Folder 1.20.

[57] Richard P.Feynman, The Character of Physical Law (London: Penguin Books, 1992).

[58] Richard P.Feynman, QED. The Strange Theory of Light and Matter (Princeton: Princeton University Press, 1985).

## *Conclusion*

A similar cult of genius has ever expanded around any other American physicist such as the one around Richard Feynman. Feynman's autobiographical writings played an essential role in the genesis of this cult of genius. In these writings, he put numerous anecdotes that circulated around his character into written form, which significantly contributed to brilliant physicists and Nobel Prize laureates lifting him up to the gallery of genius. Quite contrary to the Feynman's staging, the historical biographical case study in the first part of this essay clarified that his physical works by no means resulted free of context, but rather were the subject of a continuous thought style oriented endeavor to make his results available to the scientific community. The cult of genius around Feynman that began to grow even more after the bestowal of the Nobel Prize, increasingly concealed Feynman's foothold in the thought collective. Feynman significantly contributed to the growth of the cult of genius with his stories and autobiographical writings. In his biographical self-portrayal, he was successful in tying together his later social and scientific norms of a contradictory nature with the American myth of the ''practical man''. In public, this allowed him to appear as a typical American with extraordinary scientific capabilities that were no longer comprehendible for the average Joe or that shortly appeared to be genius.

---

[59] Richard P. Feynman, Robert B. Leighton, and Matthew L. Sands, The Feynman Lectures on Physics. 3 vols. (Redwood City, CA, Addison-Wesley, 1989).